\documentclass[aps,pra,twocolumn,showpacs,longbibliography,superscriptaddress]{revtex4-1}
\usepackage[pdftex]{graphicx}

\begin{document}

\title{Phase-Shift Plateaus in the Sagnac Effect for Matter Waves}

\author{M. C. Kandes}
\affiliation{Computational Science Research Center, San Diego State University, San Diego, CA 92182, USA}
\affiliation{Institute of Mathematical Sciences, Claremont Graduate University, Claremont, CA, 91711, USA}

\author{R. Carretero-Gonz\'alez}
\affiliation{Computational Science Research Center, San Diego State University, San Diego, CA 92182, USA}
\affiliation{Department of Mathematics and Statistics, San Diego State University, San Diego, CA 92182, USA}

\author{M. W. J. Bromley}
\homepage[]{http://www.smp.uq.edu.au/people/brom/}
\affiliation{School of Mathematics and Physics, The University of Queensland, Brisbane, QLD, 4072, Australia}
\affiliation{Computational Science Research Center, San Diego State University, San Diego, CA 92182, USA}


\date{\today}

\begin{abstract}

We simulate ultracold Sagnac atom interferometers using quantum-mechanical matter wavepackets, e.g. Bose-Einstein condensates, that counter-propagate within a rotating ring-trap.  We find that the accumulation of the relative phase difference between wavepackets, i.e. the matter wave Sagnac effect, is manifested as discrete phase jumps.  These plateaus result from three effects; that the atoms should be initially trapped at rest with respect to the rotating frame, that they counter-propagate with the same group velocities in the rotating frame, and that the imaging is performed in the rotating frame.  We show that the plateaus persist in the presence of nonlinear atom-atom interactions, and in atoms undergoing various rotations, and thus will occur during matter wavepacket experiments.  We also introduce the simplest possible Sagnac atom interferometry scheme which relies on wavepacket dispersion around a ring-trap.

\end{abstract}




\maketitle

\newpage

\begin{figure}
    \includegraphics[scale=0.40]{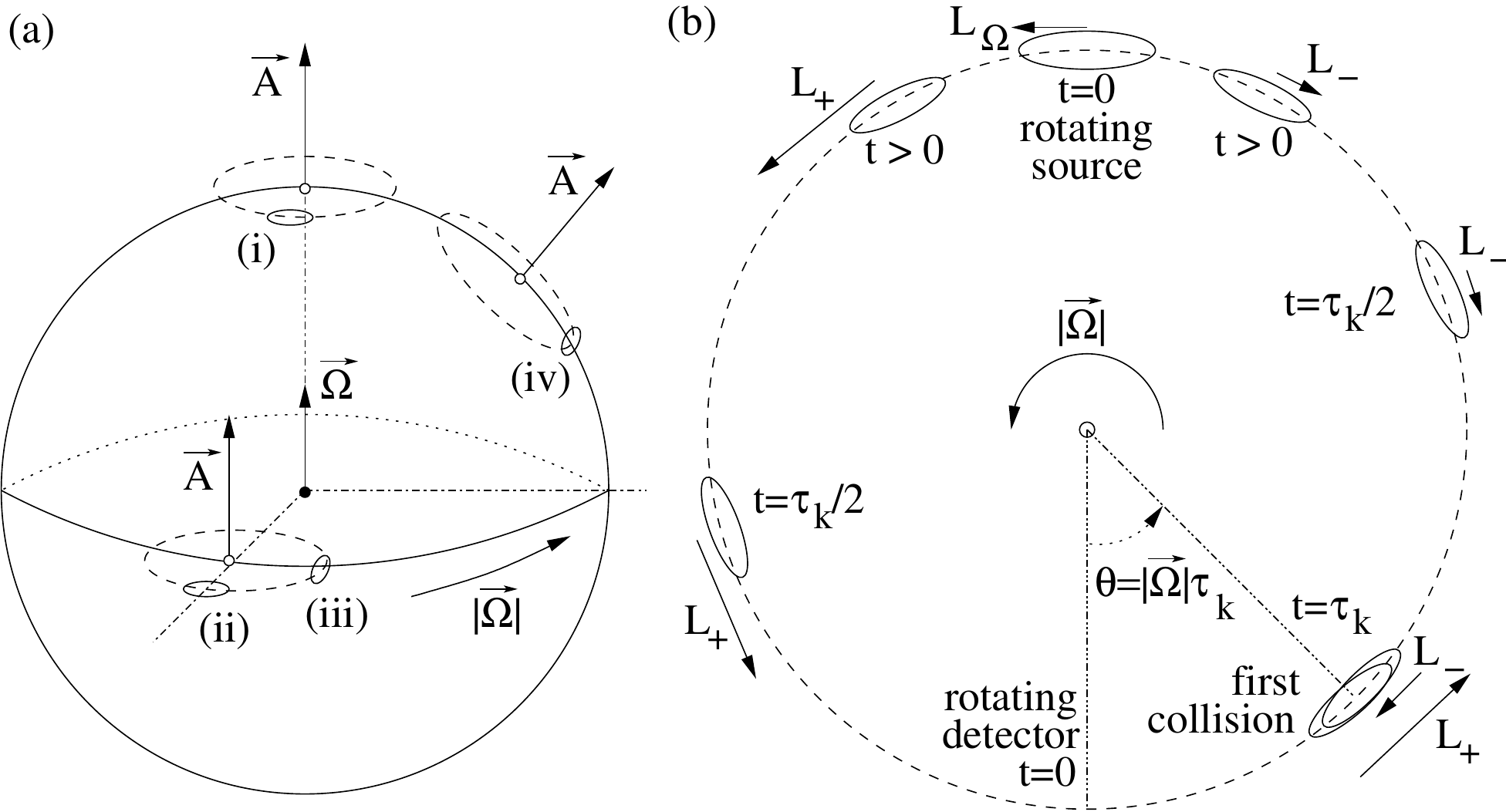}
   \caption{\label{fig:schematic}
      (a) Initial location of the various wavepackets and the areal vector
      $\vec{A}$ orientations of the ring-traps considered in this paper,
      all undergoing the same rotation rate $\vec{\Omega}$.
      (b) Schematic of our Sagnac thought experiment with localised matter
      waves as seen in the inertial (non-rotating) frame. For $t<0$ the initial wavepacket is
      localised in a rotating trap potential (i.e. with angular momentum $L_\Omega$),
      and at $t=0$ is symmetrically split into counter-propagating wavepackets in
      a ring-trap with different (group) velocities in the inertial frame.
      These collide and produce an interference pattern at $t=\tau_k$.
     }
\end{figure}

The Sagnac effect was first demonstrated experimentally for
light one hundred years ago by French physicist Georges Sagnac~\cite{sagnac13ab}
and, in recent years, atoms have begun to exhibit a rotation measurement
sensitivity able to go beyond that of light-based systems~\cite{durfee06a}.
The Sagnac effect is an interference phenomena of waves encountered in rotating
frames of reference that can be used to measure absolute rotations with respect
to an inertial frame~\cite{malykin00a}, 
such that the phase-shifts are well-known, well-tested, and are given by~\cite{cronin09a}
\begin{equation}
  \label{eqn:sagnac}
  \Delta_{\mathrm{matter}} = \frac{2m}{\hbar} \left(\vec{\Omega} \cdot \vec{A}\right) \: ,  \quad
  \Delta_{\mathrm{light}} = \frac{4\pi}{\lambda_\mathrm{light} c} \left(\vec{\Omega} \cdot \vec{A}\right) \:.
\end{equation}
where $m$ is the particle mass, $\hbar$ is the reduced Planck constant,
$c$ is the speed of light, and $\lambda_\mathrm{light}$ is the wavelength of the
light.  In both cases the shift is proportional to the magnitudes of the
areal vector $\vec{A}$ and the angular velocity vector $\vec{\Omega}$
(see Fig.~\ref{fig:schematic}(a)), however, it is (to first-order) independent
of the velocity of the rotating matter waves~\cite{varoquaux08a}.
The optical Sagnac effect has not only been used to address fundamental scientific
questions (e.g. as in the Michelson-Gale-Pearson experiment~\cite{michelson25b}), 
but it has also found technological applications.
In particular, the Sagnac effect is the fundamental basis of rotation
measurements made with fiber-optic and ring-laser gyroscopes~\cite{stedman97a},
which are currently deployed in many high-precision inertial
navigation systems.

The Sagnac effect has also been observed for matter waves, including
neutrons~\cite{werner79a},  
neutral atoms~\cite{riehle91a},
electrons~\cite{hasselbach93a} and superfluids~\cite{sato12a}. 
In this paper we ignore the (translated) advice of Malykin,
who stated ``quantum-mechanical calculations of the imaginary part of
a wave function are not at all necessary to compute the phase-shift of
de Broglie counter-propagating waves in a rotating ring
interferometer, attributable to the Sagnac effect"~\cite{malykin00a}.
Instead, we present a theoretical scheme and calculations to elucidate
how the Sagnac effect is manifested in localized matter waves.
The basic setup, schematically shown in Fig.~\ref{fig:schematic}(b),
involves trapping a BEC, splitting it into two counter-propagating
waves within a ring-shaped atom trap, and allowing them to collide.
Instead of a linear accumulation of the phase-difference with time,
we find perfectly-step-like jumps of the phase-shift between
wavepacket collisions, with the magnitude of the phase jumps being exactly
predicted by the Sagnac formula.  These plateaus will be observed in
rotating BEC experiments in the various geometries
shown in Fig.~\ref{fig:schematic}(a).

The motivation for developing Sagnac systems with atoms is
the potential improvement,
$\Delta_{\mathrm{matter}}/\Delta_{\mathrm{light}} \! \approx \! 10^{11}$
over photons (near optical wavelengths)~\cite{meystre01a}.
A state-of-the-art atom-based Sagnac interferometer is currently able to
achieve a precision comparable to ring-laser-based gyroscopes~\cite{durfee06a}.
This was based on thermal atoms in a non-portable $2$~m long vacuum chamber with
an enclosed area of $\approx 30$~mm$^2$, i.e. an effective circle of radius $3$~mm.

While there are competing technologies developing
compact sensors~\cite{sato12a}, ultracold gaseous atoms
in a vacuum chamber offer a flexible scaling of the
geometry~\cite{schnelle08a}, and thus could explore
systematics on the fly.
A BEC-based rotation sensor would enable a relatively
compact sensor with long experiment times~\cite{cronin09a}
and the use of common mode noise rejection by making the atoms enclose
loops, e.g. using guided BEC configurations~\cite{gupta05a}
and portable atom chips~\cite{wang05a}.  
Such a sensor could be used on Earth~\cite{cronin09a},
but it would also be feasible
to send it into space~\cite{muntiga13a},
in the search for physics beyond the standard model~\cite{chung09a}.
The best attempt so far at the measurement of the rotation of the Earth in
a BEC experiment~\cite{burke09a} enclosed areas of up to $0.1$ mm$^2$,
i.e.  an effective radius of $200$ $\mu$m (at the limit for being able
to observe the Earth's rotation), but was unable to do so due to
technical challenges.

Apart from schemes where a localised BEC is split and traverses a
guided path around the interferometer~\cite{burke09a}, there have
been schemes proposed to use the interference due to the
counter-rotation of BEC modes~\cite{halkyard10a}
which are yet to be realised experimentally.
There has also been recent experimental progress in this direction with a
Sagnac measurement using sound-waves in a BEC in a ring-trap~\cite{marti12a}.
The key idea that differentiates such systems from the present paper
is that their propagating BEC modes fill up the ring-trap and hence
they experience a linear accumulation of the phase shift,
i.e. $\vec{A} = \int_0^t \frac{d\vec{A}}{dt^\prime} dt^\prime$),
that can be extracted at any measurement time.
In contrast, for wavepackets we need to wait until a collision.

\section{1-D thought experiment}

\begin{figure}
   \includegraphics[scale=0.8]{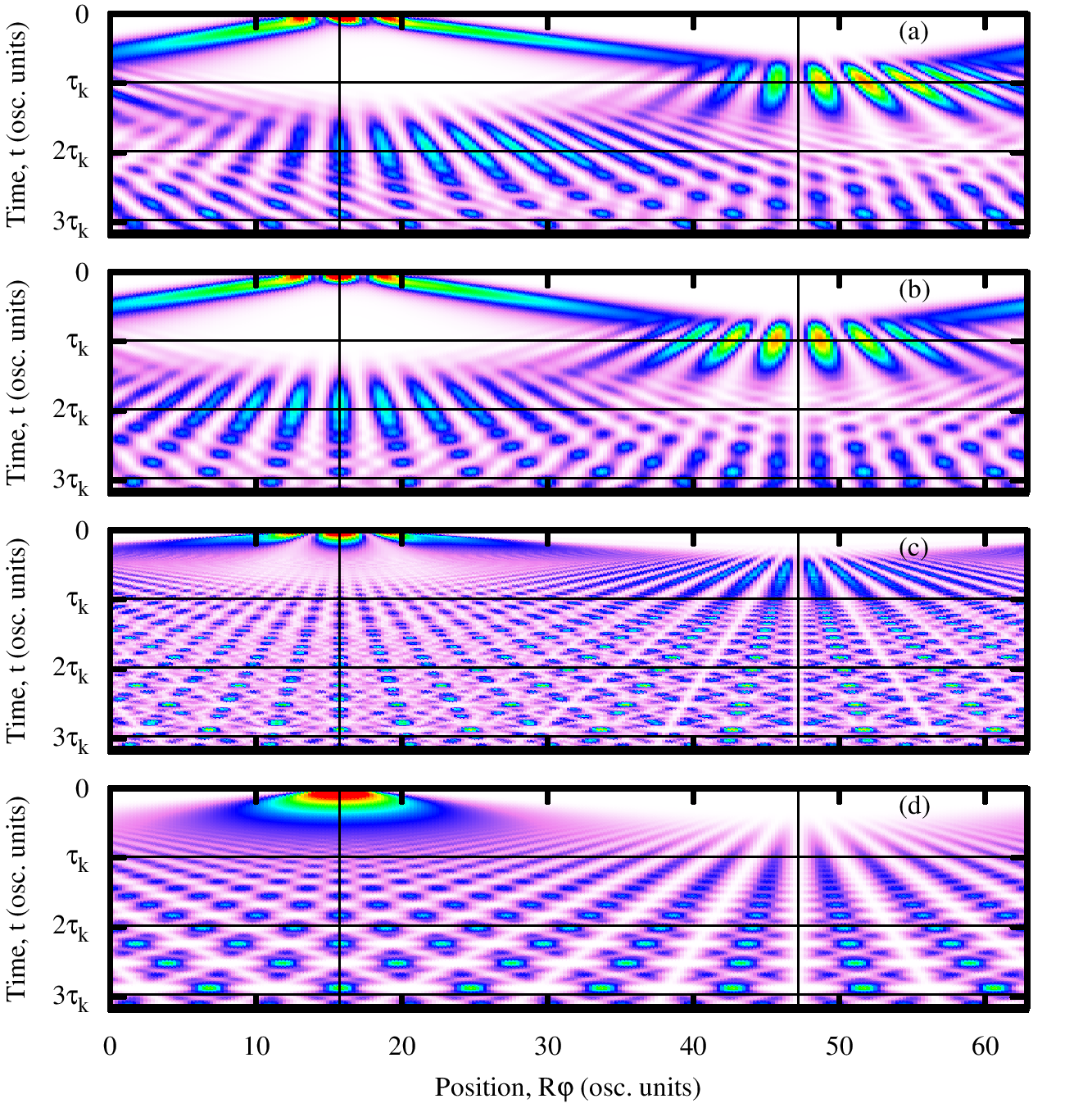}
   \caption{\label{fig:spacetime-density} (colour online) Spacetime plot of the probability
      density for counter-propagating wavepackets (as viewed in the rotating frame)
      initially trapped with $\omega_\phi = 0.1$
      in an $\Omega=0.0150$ rotating 1-D ring of radius $R=10$, with
      (a) $L_0 = 0$ and $g_1=0$,
      (b) $L_0=\Omega R^2$ and $g_1=0$, and
      (c) $L_0=\Omega R^2$ and $g_1=10$.
      All units given in oscillator units.
      Symmetric angular kicks have been applied at $t=0$ with
      $L_k = 10$.
      Plot (d) shows a dispersion-based ($L_k=0$) atom interferometer with
      $\omega_\phi = 1$, $L_0=\Omega R^2$, $g_1=0$.
      The highest probability densities are coloured and the
      lowest densities are white.
      The vertical black lines indicate the 
      azimuthal positions $\frac12 \pi R\approx15.7$ and $\frac32 \pi R \approx47.1$ on 
      the ring. The horizontal black lines indicate the
      times, $\tau_k = 10\pi$ osc. units, at which two symmetrically counter-propagating
      classical particles with angular momentum $L_k=10$ would undergo 1, 2, 3 collisions in the ring.
     }
\end{figure}

We start with the geometry of case-(i) of
Fig.~\ref{fig:schematic}(a) to walk through the
thought experiment of Fig.~\ref{fig:schematic}(b).
The rotation here has cylindrical ($\rho,\phi,z$) symmetry about the axis
passing through the ring centre and normal to the plane in which the ring lies.
The ring-trap potential we choose has harmonic confinement in the
transverse ($\rho$ and $z$) directions, with
$V_{\mathrm{ring}}(\rho,\phi,z) = \frac12 \omega_\perp^2 (z^2 + (\rho-R)^2)$
with oscillator units used throughout this paper (see Appendix~\ref{app:model}).
Here we approximate this setup as a 1-D ring since the rotation
term in the Hamiltonian is $-\vec{\Omega} \cdot \vec{L} = -\Omega \hat{L}_z = i \Omega \frac{\partial\:}{\partial \phi}$,
and so we choose ans{\"a}tze
$\Psi(\rho,\phi,z,t) = \psi_\rho \psi_z \psi(\phi,t)$.
The $\psi_\rho$ and $\psi_z$ are frozen-in-time Gaussians based
on atom traps of frequency $\omega_\perp = 1$, with all of the
1-D ring dynamics then described by $\psi(\phi,t)$.
Four 1-D calculations are shown in the
space-time plots of Fig.~\ref{fig:spacetime-density},
which depict the time-dependent probability density for atom/s
in a ring-trap of radius $R=10$ which is rapidly rotating
at $\Omega=0.0150$ (in osc. units, e.g. at $1.5$ Hz).
The first two plots involve (linear) Schr{\"o}dinger physics with
a single atom, $N=1$ (see Appendix~\ref{app:model}; thus $g_3=0$),
the third plot has the nonlinear interactions through a strong
1-D nonlinear coupling constant $g_1 = g_3/(2\pi\beta_\perp^2) \equiv 10$
(in osc. units, e.g. $\approx 1000$ $^{87}$Rb atoms).
The first three plots in Fig.~\ref{fig:spacetime-density} highlight the four
operations (hereafter numbered I, II, III and IV)
that our thought experiment demands, which we now discuss sequentially.

(I) The wavepacket requires an initial state.  The wavefunction is
initially loaded within a ring of radius $R$ and
localised for $t<0$ at an angle $\phi_0$ by an additional
potential,
$V_{\mathrm{trap}}(\phi,t<0) = \frac12 \omega_\phi^2 R^2(\phi-\phi_0)^2$,
by, for example, lingering a laser-induced ring potential longer
around $\phi_0$~\cite{schnelle08a}.
This breaks the rotational invariance of the Hamiltonian,
with solutions
$\psi_0(\phi) = C \exp(-\frac12 \omega_\phi R^2(\phi-\phi_0)^2) \exp(i L_0 (\phi-\phi_0))$.
Here $C$ is the normalisation, while $L_0$ is the initial angular momentum
of the wavepacket which depends on what reference frame one 
considers the atom/s are initially at rest with respect to.
For example, Fig.~\ref{fig:spacetime-density}(a) is a simulation with
the atom initially at rest with respect to the inertial frame (i.e. $L_0=0$).
In contrast, however, if the atom is at rest with respect to
the co-rotating frame of the ring, as seen in
Fig.~\ref{fig:spacetime-density}(b), then $L_0=L_\Omega=\Omega R^2$.
However, at what speed should the atom trap itself be rotating?
Since in a general Sagnac experiment the apparatus will
be rotating, e.g. on Earth, then the atom trap must
be stationary with respect to the rotation. Thus
Fig.~\ref{fig:spacetime-density}(a) is artificial, whilst
Figs.~\ref{fig:spacetime-density}(b,c) are realistic situations.

(II) The wavepacket is then split at $t=0$.
This is required to mimic a near-ideal diffraction of an
atom using lasers~\cite{wu05a}.
A linear superposition of two counter-propagating wavepackets,
%
   $\psi(\phi,0) = A \ \psi_0(\phi) \ e^{iL_A(\phi-\phi_0)}
                + B \ \psi_0(\phi) \ e^{-iL_B(\phi-\phi_0)}$,
%
serves as the state at $t=0$.
The $A$ and $B$ are normalisation constants, controlling the
split between the left and the right (here we assume 50\%/50\% splits).
The angular momenta imparted to the atom/s by the laser are $L_A$ and $L_B$.
However, it is crucial to discuss what $L_A$ and $L_B$ should be,
i.e. what (group) velocity should the atoms traverse the ring with?
The textbook treatment of this
(see, for example, Chapter 11.4 in Ref.~\cite{foot05a})
says that the atoms traverse with the same (group) velocity in the inertial frame
and the correct result for the matter wave Sagnac shift is thus derived.
This, however, is not the natural answer for atoms.  Since, in general,
the lasers and mirrors performing the split are also co-rotating
with our initial atom trap ($L_0 = L_\Omega$) then they are
(approximately) travelling with the same tangential velocity.
Upon being kicked atoms should thus obey a Galilean-type
addition/subtraction of velocities, i.e. a matter wave
equivalent of the Ritz ballistic hypothesis~\cite{malykin10a},
given low velocities.
The atom kicks applied must then be equal and opposite in direction
in the \textit{rotating reference frame} ($L_k = L_A = L_B$),
which gives different velocities in the inertial frame
(i.e. in Fig.~\ref{fig:schematic}(b) we have $L_\pm = L_\Omega \pm L_k$).
In the rotating reference frame, our scheme results in symmetric
wavepacket `jets' for $0 \! < \! t \! < \! 20$ as seen in Figs.~\ref{fig:spacetime-density}(b,c).
The asymmetry in Fig.~\ref{fig:spacetime-density}(a) is due to the
atom trap being stationary in the inertial frame ($L_0 = 0$) and thus
symmetric $L_k$ kicks result in differing group velocities
in the rotating frame.

(III) The wavepackets evolve for some time $t$ in a ring-trap
while the relative phase-shift between them accumulates with time,
which is called the Sagnac effect
[see Appendix~\ref{app:derivsagnac} for our derivation of Eq.~(\ref{eqn:sagnac})].
For the systems in Fig.~\ref{fig:spacetime-density},
the Sagnac shift is $\Delta = 2\times 0.0150\times \pi 10^2 = 3\pi$
which is seen as an interference pattern that runs down a density minima on the
opposite side of the ring (at $\phi = \frac32 \pi$) during the first
collision near $t = \tau_k = \pi R^2/L_k = 31.4$, which
is the time that classical particles would collide.
By the time of the second collision ($t = 2\tau_k = 62.8$), back at
the starting location ($\phi = \frac12 \pi$), the interference patterns
run down the maxima with respect to time, indicating $\Delta = 6\pi$.
The phase of the interference in Fig.~\ref{fig:spacetime-density}(b)
does not appear to change in time during each collision.
This same behaviour occurs in Fig.~\ref{fig:spacetime-density}(c)
despite the nonlinear dispersion of the $g_1=10$ wavepacket
(as seen for $0 \! < \! t \! < \! 20$), which is enhanced by using
the same $\psi_0$ initial state for the calculations in both
Figs.~\ref{fig:spacetime-density}(b) and (c).


(IV) The measurement will always be performed in the rotating frame,
i.e. our detector is co-rotating with the atoms.  We `image' the
probability density at a particular time, producing one of the
slices in time which are used to build Fig.~\ref{fig:spacetime-density}.
We take this slice and extract a single phase-shift with a
Fourier transform at the dominant frequency of the collision
interference (here at wavenumber $2k_\phi \equiv 2L_k/R$; see Appendix~\ref{app:phaseshifts}).
In Fig.~\ref{fig:phase-shift-vs-time}, the phase-shifts extracted
as a function of time are shown in simulations using
four speeds of rotation.
These are presented with the same cases seen in
Fig.~\ref{fig:spacetime-density}, as well as the same kicks $L_k=10$,
however, slower rotation rates were chosen here so that the phase shifts
stay below $\pi$ for clarity.
%
\begin{figure}
    \includegraphics[scale=0.600]{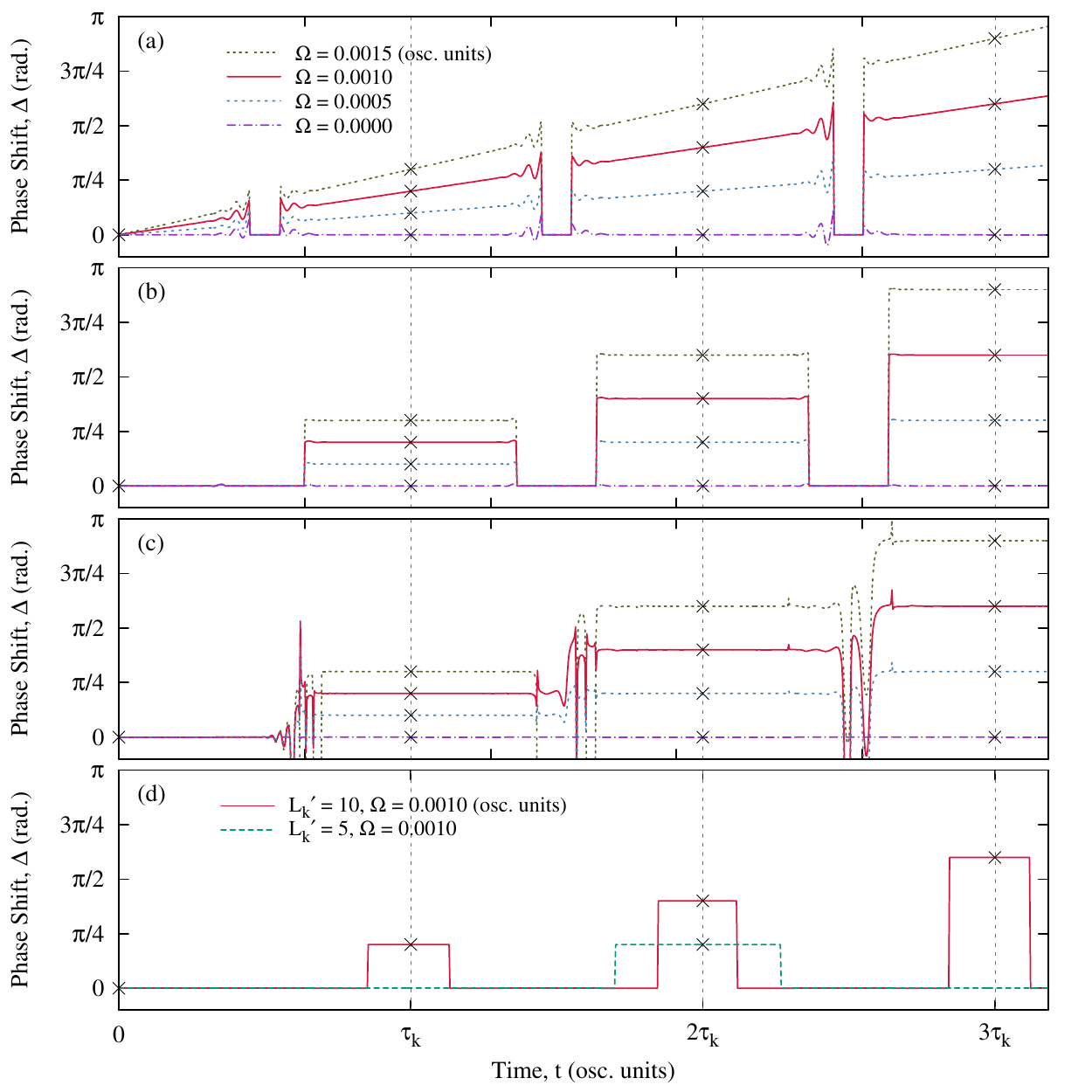}
    \caption{\label{fig:phase-shift-vs-time} 
       (colour online) Sagnac phase-shifts extracted
       from 1-D calculations as a function of time
       (in $\tau_k = 10\pi$ osc. units), for counter-propagating
       single-particle wavepackets split with a kick $L_k = 10$ where
       (a) $L_0=0$, $g_1=0$, (b) $L_0=\Omega R^2$, $g_1=0$,
       and (c) $L_0=\Omega R^2$, $g_1=10$.
       The different curves correspond to $R=10$ rings rotating
       at various rates $\Omega$ as indicated in the top panel,
       all extracted using Fourier components with $2k_\phi = 20$.
       Plot (d) shows a single dispersion-based atom interferometer for
       one rotation rate $\Omega = 0.0010$ with
       $\omega_\phi = 1$, $L_0=\Omega R^2$, $g_1=0$, and no kick ($L_k=0$).
       Plot (d) shows two sets of phase-shifts that are simultaneously extracted
       from each calculation using effective momenta $L^\prime_k = 5$ and $10$.
       We set $\Delta(t)=0$ when there is not enough signal (see Appendix~\ref{app:phaseshifts}).
       The $\times$-symbols denote the predicted Sagnac shifts using Eq.~(\ref{eqn:sagnac}).
       }
\end{figure}

For the artificial case of Fig.~\ref{fig:spacetime-density}(a) we find that
the time-dependent phase-shift accumulates linearly with time, exactly
at the rate predicted by Eq.~(\ref{eqn:sagnac}) as one might expect from
previous studies of Sagnac atom interferometry~\cite{halkyard10a,marti12a}.
For the case depicted in Fig.~\ref{fig:spacetime-density}(b) we find that the
time-dependent phase-shift response undergoes discrete phase jumps
between the classical collision times at multiples of $t = \tau_k \approx 31.4$.
We also observe that the magnitude of the observed phase jumps are
dependent only upon the magnitude of the angular velocity of the ring ($\Omega$)
and not dependent on the kick ($L_k$).
Performing the same analysis for the nonlinear case, as per
Fig.~\ref{fig:spacetime-density}(c), we continue to observe perfectly
flat plateaus.  Essentially, despite the extra dispersion, this shows that the
Sagnac effect remains independent of the (group) velocity
of the counter-propagating atoms (see Ref.~\cite{malykin00a}).
Without any dispersion, the interference fringes during the collisions
are seen to be perfectly stationary in the rotating frame as a consequence
of having the split wavepackets with $L_0 = L_\Omega$
[see Appendix~\ref{app:derivplateaus} for a derivation].

\section{Dispersion-based Sagnac interferometer}

We now introduce a simplified, dispersion-based, Sagnac atom interferometer
by removing step (II) from the system, i.e. the self-induced expansion
of a single wavepacket will also realise an interferometer~\cite{chen03a}.
This is seen in 1-D calculations in Fig.~\ref{fig:spacetime-density}(d),
where we use a tighter initial trap, with $\omega_\phi = 1$, to
enhance the dispersion.  The interference pattern 
is again seen to exhibit a constant ($3\pi$) phase-shift on
the opposite side of the ring from the release.
Since no splitting kick is introduced ($L_k=0$), there is no
dominant frequency to extract phase-shifts at.
Nonetheless, we are able to observe phase-shift plateaus due to
the spectrum of frequencies that compose a wavepacket, with
each frequency observed in the interference (see Fig.~\ref{fig:spacetime-density}(d)).
The phase-shifts are shown in Fig.~\ref{fig:phase-shift-vs-time}(d)
for $\Omega = 0.0010$, where we have simultaneously extracted the
phase-shifts from the same interference patterns using two
effective $L^\prime_k=5$ and $L^\prime_k=10$ momenta 
in the Fourier analysis.  As each $L^\prime_k$ momentum
component propagates at a different (group) velocity around
the ring, each has its own particular collision time 
($\tau_{k^\prime}$), around which one is able to image the
atoms and extract the Sagnac phase-shift.

\begin{figure}
   \includegraphics[scale=0.675]{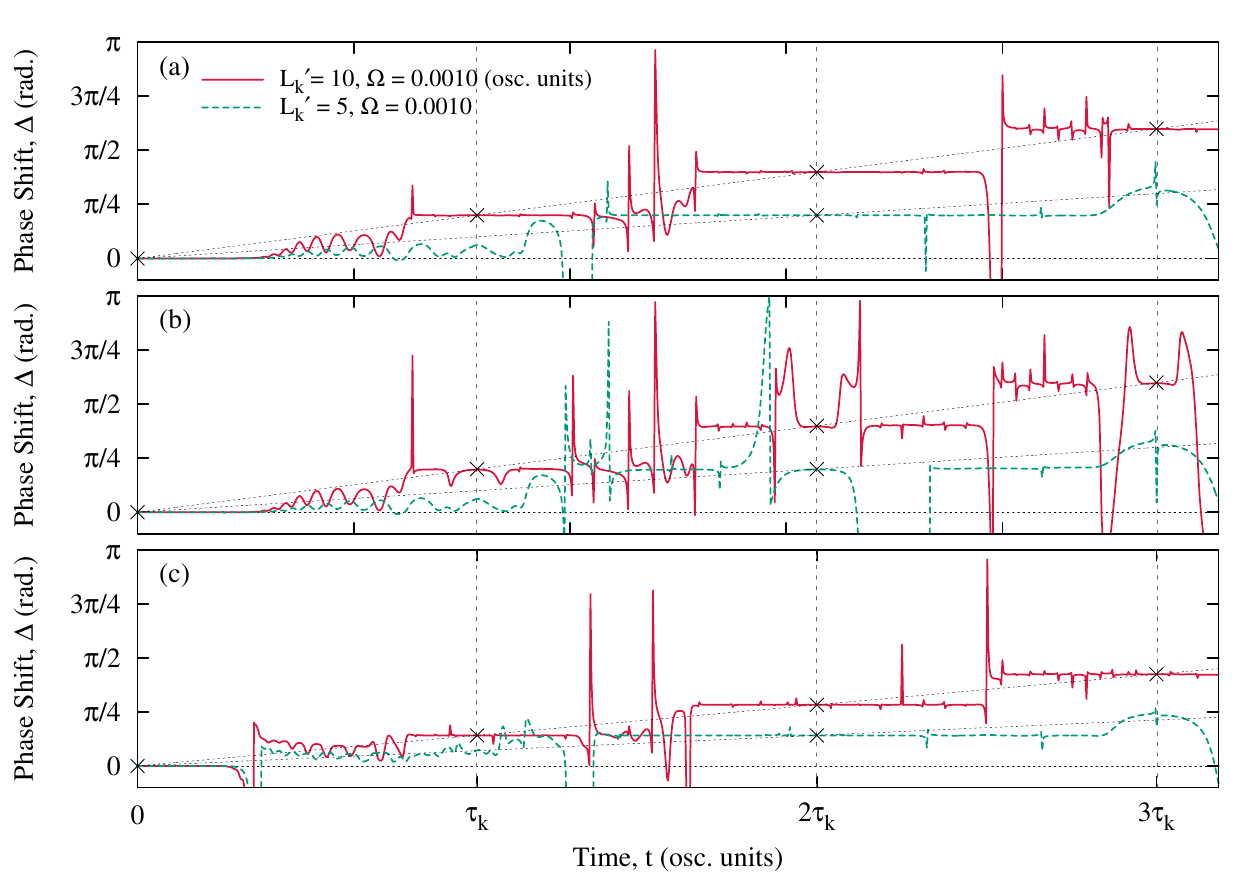}
   \caption{\label{fig:phase2d3d} 
       (colour online) Sagnac phase shift extracted as a function of time
       (in $\tau_k = 10\pi$ osc. units) for three ($L_\phi=0$) dispersion-based
       atom interferometers with the same rotation rate of $|\vec{\Omega}| = 0.0010$.
       We use a ring with $R=10$ initially localising the wavepacket in an
       angular trap, using $\omega_\phi=1$.
       The (a) and (b) panels are from 2-D calculations with
       $\phi_0 = 0$ and $\frac12 \pi$ respectively, corresponding
       to cases (ii) and (iii) of Fig.~\ref{fig:schematic}(a), while
       panel (c) is from a 3-D calculation as per case (iv) of
       Fig.~\ref{fig:schematic}(a) with the initial dispersion moving
       into/away from the direction of rotation.
       Two sets of phase-shifts are simultaneously extracted 
       from each calculation using effective momenta $L^\prime_k = 5$ and $10$.
       The inclined lines depict the constant rate of Sagnac
       accumulation required to achieve a phase-shift given
       by Eq.~(\ref{eqn:sagnac}) per `collision' (denoted by $\times$-symbols).
     }
\end{figure}

We now remove the $\psi_\rho$ confinement and show
that plateaus persist in 2-D calculations (neglecting gravity),
corresponding to cases (ii) and (iii) of Fig.~\ref{fig:schematic}(a).
These require the ans{\"a}tze $\Psi(x,y,z,t) \approx \psi_z \psi(x,y,t)$,
given a ring-trap
$V_{\mathrm{ring}}(\rho,\phi,z) = \frac12 (\omega_\perp^2 z^2 + \omega_\rho(\rho-R)^2)$
and a localising initial potential
$V_{\mathrm{trap}}(\phi,t<0) = \frac12 \omega_\phi^2(\phi-\phi_0)^2$.
We choose a $R=10$ ring with confinement strengths:
$\omega_\perp = 1$ (which sets the length scale),
$\omega_\rho = 10$ (for tight radial confinement),
and $\omega_\phi = 1$ (for angular localisation).
We break the rotational symmetry by placing the rotation
`off-axis' at $(x_\Omega,y_\Omega) = (-40,0)$ (see Appendix~\ref{app:model}),
with the ring centred at $(x,y)=(0,0)$ in Cartesian-based calculations.
An initial state as per case (ii),
located near $(x_0,y_0)=(10,0)$, has an initial phase gradient
across the wavefunction that is similar to the 1-D calculations
in Fig.~\ref{fig:spacetime-density}(b).
As the wavefunction disperses with time, we integrate the
2-D density onto a 1-D ring and extract the phase-shifts
(at two different effective momenta $L^\prime_k = 5$ and $10$),
which clearly exhibits plateaus as shown in Fig.~\ref{fig:phase2d3d}(a).
We take this further in Fig.~\ref{fig:phase2d3d}(b),
which realises case (iii) in Fig.~\ref{fig:schematic}(a),
with the initial state localised with $\phi_0 = \frac12 \pi$,
i.e. at $(x_0,y_0)=(0,10)$.
This is a different thought experiment to that setup
in Fig.~\ref{fig:schematic}(b). The wavepacket dispersion
is initially along the $x$-axis, i.e. into/away from the
center of the rotation (effectively $L_0=0$ around the
direction of the ring).  The dynamics are complicated, however,
the phase-shifts do briefly touch into plateaus around the
`collision' points.  Note that for case (iii)-systems, as the rotation
rate is increased, the plateaus become increasingly distorted
due to the dynamics induced by the angular asymmetry of the initial state
(due to the centrifugal force), which is not experienced for
a case (i) or (ii) system [see Appendix~\ref{app:videos} and Ancillary videos].

The final case that we consider
is the 3-D calculation of Fig.~\ref{fig:schematic}(a) case (iv).
We choose to align the rotating frame such that the $x>0$ axis aligns
to the local north, $z>0$ aligns away from the origin of the rotation
(at $(x_\Omega,y_\Omega,z_\Omega) = (0,0,-40)$), while $y>0$ aligns
westward (away from the rotation). 
The angular velocity is chosen as
$\vec{\Omega} = \Omega\:[\frac{1}{\sqrt{2}},0,\frac{1}{\sqrt{2}}]$
(with $\Omega = 0.0010$).
Our $R=10$ ring-trap potential, again with 
$\omega_\perp = 1$ and $\omega_\rho = 10$,
has $\vec{A}=[0,0,\pi 10^2]$. Our angular harmonic trap with
$\omega_\phi = 1$ and $\phi_0=\frac32 \pi$ localises the wave for
$t<0$ near $(x_0,y_0,z_0)=(-10,0,0)$.
After the release into the 3-D ring at $t=0$, in Fig.~\ref{fig:phase2d3d}(c),
we again see plateaus around the `collision' times of the different
(angular) momentum components, of step-sizes that are in perfect
agreement with Eq.~(\ref{eqn:sagnac}), given the reduced product
$\vec{\Omega} \cdot \vec{A} = \Omega A / \sqrt{2}$.

In conclusion, we have described how the Sagnac phase shift
accumulates in time as observed through a set of numerical simulations
modeling the dynamics of a matter wave Sagnac interferometer.
In the end, Malykin was correct~\cite{malykin00a}, and one can
mostly ignore the complex nature of the Sagnac effected matter wave.
However, by 
considering a thought experiment, where the wavepackets are initially
at rest with respect to the rotating frame, we have found that
the phase shift accumulates in apparent discrete phase jumps,
and proposed a simple dispersion-based Sagnac atom interferometer.
One advantage of these observations for experimentalists is that
the phase shift is insensitive to the exact time that the
measurement is performed (although maximum contrast will be
at the peak of the collision).
This measurement insensitivity will also likely apply if the
experiment involves a recombination laser pulse followed by a
time-of-flight expansion to extract the number of atoms in the
two output ports~\cite{burke09a}.
Future work will involve examining the robustness of the plateaus
to further perturbations than in the systems examined here,
and the inclusion of finite-temperature effects~\cite{chen03a}.
Our results may guide the design of high-precision rotation sensors,
especially for measurements aimed at studying rotational phenomena occurring
at low angular velocities~\cite{lantz09a}.
Progress in these regimes is likely to come from the continued development
of neutral atom Sagnac interferometers, particularly involving
number squeezed states~\cite{lucke11a}. \\

\begin{acknowledgments}

This research was supported by a graduate student teaching 
assistantship in the Department of Physics at San Diego State University,
Cymer Incorporated and the San Diego Chapter of the ARCS Foundation,
the Inamori Foundation, as well as the Australian Research Council
through a Future Fellowship (FT100100905). Thanks to B.M. Fahy for
3-D parallel code development and S.A. Haine for instructive conversations. \\

\end{acknowledgments}

%
\appendix

\section{Modelling the atoms}
\label{app:model}

To model a rotating ultracold Bose gas we rely on the
nonlinear Schr{\"o}dinger Equation (NLSE)~\cite{lieb06a},
\begin{equation}
 \label{eqn:rnlse}
     i \hbar \frac{\partial \psi}{\partial t}
  = \left[-\frac{\hbar^2}{2m} \nabla^2 + V_r({\bf r},t) - \vec{\Omega} \cdot \vec{L}
        + g_3 |\psi|^2 \right] \psi \: .
   \end{equation}
As per (number conserving) theory~\cite{esry97a},
this describes the evolution of a single atom, $\int |\psi|^2 dr^3 = 1$,
while the nonlinear coupling constant, $g_3=4\pi \hbar^2 a_{s}(N\!-\!1)/m$,
characterises the short-range pairwise interactions between it and the
other $(N\!-\!1)$ bosons in the gas.
This depends on the {\em s}-wave atom-atom scattering length,
$a_{s}$, of two interacting bosons.  Note that alternative treatments
give $g_3 \propto N$, and then the NLSE is known as the
Gross-Pitaevskii equation (GPE)~\cite{esry97a,lieb06a}. 
The rotating configuration of the potential, $V_r({\bf r},t)$, here will generally
be time-independent in some reference frame,
but we do abruptly switch between potentials at $t=0$.
The transformation to the rotating reference frame is achieved with
angular momentum operators $\vec{L} = [\hat{L}_x,\hat{L}_y,\hat{L}_z]$.
Despite its simplicity, Eq.~(\ref{eqn:rnlse}) accounts for the
inertial forces that may act on the atom in the rotating reference
frame --- namely, the centrifugal force, the Coriolis force
(and, if $\vec{\Omega}$ was time-dependent, the Euler force).

We use the Crank-Nicolson finite-differences method,
employing split-operator algorithms~\cite{bromley04a} for 2-D and 3-D parallelisation.
For 2-D and 3-D calculations we centre the ring at $(x,y)=(0,0)$
on a Cartesian-grid at $(x,y)=(0,0)$ with generalised rotation operators e.g.
$-\Omega \hat{L}_z = i\hbar \Omega \left((x-x_\Omega)\frac{\partial \:}{\partial y} - (y-y_\Omega)\frac{\partial \:}{\partial x}\right)$
where $(x_\Omega,y_\Omega)$ is the origin of the rotation.
We imprint a Gaussian-based initial wavefunction with the
correct phase gradient to match the rotation,
and use imaginary time propagation 
to determine the initial state of the 2-D and 3-D systems
that is stationary in the rotating frame.

We rescale Eq.~(\ref{eqn:rnlse}) in harmonic oscillator units,
which rescales energies, lengths and times by
$\hbar\omega_\perp$, $\beta_\perp=\sqrt{\hbar/m\omega_\perp}$
and $\omega_\perp^{-1}$, respectively,
where $\omega_\perp$ is some natural angular frequency
of the system.  For example, for a gas of $^{87}$Rb
with $\omega_\perp=2\pi \times 100$ Hz then, for our results presented
in this paper, lengths are in units of $\beta_\perp \approx 1.1$ $\mu$m and
time is in units of $1.6$ ms.
The rotation of the Earth, for example, is then $\Omega \approx 10^{-7}$ osc. units.

\section{Extracting phase-shifts}
\label{app:phaseshifts}

Our phase-shifts are obtained from the density snapshots with a
Fourier transform algorithm~\cite{goldberg01a}. This algorithm
numerically computes the phase-shift by approximating it as
\begin{equation}
   \label{eqn:fourier-transform-algorithm}
   \Delta(t) \approx \tan^{-1}\left\{\mathcal{F}_{2k_\phi}\left[|\psi(\phi,t)|^2\right]\right\} .
\end{equation}
The $\mathcal{F}$ is the Fourier transform of the atom/s probability
density computed at the dominant frequency of the interference,
here twice the wavenumber, $k_\phi$, of the colliding modes
($2k_\phi \equiv 2L_k/R$ in osc. units).
We use a numerical threshold such that $\Delta=0$ for
$|\mathcal{F}_{2k_\phi}|<10^{-10}$.
Note that we `unwrap' $\Delta(t)$ by adding/subtracting appropriate multiples
of $\pi$ (e.g. along a plateau of value $\frac12 \pi$ occasionally
the algorithm returns $-\frac32 \pi$).

\section{Derivation of Sagnac phase-shift}
\label{app:derivsagnac}

We present here an illuminating derivation of the matter-wave
Sagnac formula of Eqn.~(1) of our paper, based on the thought
experiment as presented in our paper in Fig.~(1).  This gives the
same answer as generalised treatments of the Sagnac effect as seen via
semi-classical methods \cite{hasselbach93a,meystre01a,varoquaux08a}.
It is also more illustrative than derivations based on
straightforward substitutions to convert the optical Sagnac effect
into a matter wave expression \cite{malykin00a}.
Our thought experiment is different to a textbook
treatment of the problem \cite{foot05a},
and in our proof we utilise the group and phase
velocities in the inertial frame to demonstrate the
Sagnac phase accumulation.

We begin our derivation by assuming a matter wave packet is initially 
localised in a potential undergoing uniform circular motion when viewed 
from an inertial frame of reference, with the potential located at a radius
$R$ from the axis of rotation and traversing its circular path with an 
angular velocity $\Omega$, e.g. see Fig.~1(a). As a result of being trapped
in this rotating potential, in the inertial frame the wave packet acquires
an initial group velocity and wavenumber (tangent to the circular motion) of
\begin{equation}
   v_{g_0} = R \Omega \quad , \quad k_0 = \frac{m R \Omega}{\hbar} \;.
\end{equation}
This initial wave packet is then split into a superposition of two 
counter-propagating wave packets that traverse the ring-trap, which
also has radius $R$. The out-going wave packets have group velocities
\begin{equation}
   v_{g_{\pm}} = \frac{\hbar k_{\pm}}{m} \; ,
\end{equation}
in the inertial frame, where $\hbar$ is the reduced Planck constant,
and $m$ is the particle mass, such that
\begin{equation}
   k_{\pm} = \left(k_0 \pm k\right)  \; ,
\end{equation}
are the wavenumbers of the out-going wave packets given that 
$k$ is the wavenumber symmetrically imparted to the wave packets
during the splitting process.  Assuming that the wave packets are
free particles as they traverse the ring-trap in opposite directions,
then they have phase velocities in the inertial frame and wavelengths of
\begin{equation}
     v_{p_{\pm}} = \frac{v_{g_{\pm}}}{2} \quad , \quad
    \lambda_{\pm} = \frac{2\pi}{k_0 \pm k} \; .
\end{equation}

It then follows straightforwardly that the phase of each wave packet 
advances in time as
\begin{equation}
   \varphi_{\pm}(t) = 2\pi n(t) = \frac{2\pi v_{p_{\pm}} t}{\lambda_{\pm}} \;,
\end{equation}
where $n(t)$ is the number of wavelengths the phase has shifted relative
to each wave packets peak.  Thus, the relative phase difference between the two
counter-propagating wave packets is given by
\begin{equation}
      \Delta(t)
   = \varphi_{+}(t)-\varphi_{-}(t)
   = 2\pi\left(\frac{v_{p_+}}{\lambda_{+}}-\frac{v_{p_-}}{\lambda_{-}}\right) t
   = 2R\Omega k t \; ,
\end{equation}
i.e. the phase difference accumulates linearly with time.
If we now compute the relative phase difference at the time when
the peaks of the wave packets would first coincide in the ring-trap, i.e.
\begin{equation}
   \tau_k = \frac{\pi m R}{\hbar k},
\end{equation}
(which is also when two classical particles would collide), we finally find that 
\begin{equation}
   \Delta(t=\tau_k) = \frac{2 m}{\hbar} \Omega A,
\end{equation}
where $A = \pi R^2$. This is the expected Sagnac phase-shift.

\section{Derivation of motion of fringes}
\label{app:derivplateaus}

Next, we give a simple analysis showing that the interference of the
counter-propagating wave packets produces an interference pattern that 
rotates with the group velocity of the initially trapped wave packet, 
$v_{g_0}$, in the inertial frame. Thus, when viewed in the rotating 
frame, the fringes appear stationary.

Each wave packet is assumed to counter-propagate about the ring-trap 
as a time-dependent, free-particle Gaussian wave packet of the form
\begin{equation}
   \psi_{\pm}(x,t)=Ce^{ik_{\pm}\left(x-v_{p_{\pm}}t\right)}
   e^{-\left(x-v_{g_{\pm}}t\right)^2/2\sigma^2},
\end{equation}
where $C$ is some normalization constant and $\sigma$ is the width of the
wave packets. In general, both $C$ and $\sigma$ are time-dependent,
and have complex components.  To obtain our simple proof, however,
we assume they are fixed and real, effectively ignoring the effects of
dispersion on the interference (as seen in Figs.~2(b,c)).

Computing the probability density of a superposition of these 
analytic wave packets, we find
\begin{widetext}
\begin{eqnarray}
   &   & |\psi_+(x,t) + \psi_-(x,t)|^2   \\
   & = & |\psi_+(x,t)|^2 + |\psi_-(x,t)|^2 +
          \psi_+^*(x,t)\psi_-(x,t) + \psi_-^*(x,t)\psi_+(x,t) \nonumber \\
   & = & |C|^2 \Big(e^{-\left(x-v_{g_{+}}t\right)^2/\sigma^2} +
               e^{-\left(x-v_{g_{-}}t\right)^2/\sigma^2} + \nonumber \\
   &   &       e^{-\left(x-v_{g_{+}}t\right)^2/2\sigma^2}
               e^{-\left(x-v_{g_{-}}t\right)^2/2\sigma^2} 
         \left[e^{ik_{+}\left(x-v_{p_{+}}t\right)-ik_{-}\left(x-v_{p_{-}}t\right)}+e^{-ik_{+}\left(x-v_{p_{+}}t\right)+ik_{-}\left(x-v_{p_{-}}t\right)}\right] \Big) \nonumber \\
   & = & |C|^2 \Big(e^{-\left(x-v_{g_{+}}t\right)^2/\sigma^2} + 
                    e^{-\left(x-v_{g_{-}}t\right)^2/\sigma^2} + 
         2 e^{-\left(x-v_{g_{+}}t\right)^2/2\sigma^2}
           e^{-\left(x-v_{g_{-}}t\right)^2/2\sigma^2}
   \cos\left[k_+\left(x-v_{p_+}t\right)-k_-\left(x-v_{p_-}t\right)\right] \Big)\nonumber \\
   & = & |C|^2 \Big(e^{-\left(x-v_{g_{+}}t\right)^2/\sigma^2} +
                    e^{-\left(x-v_{g_{-}}t\right)^2/\sigma^2} + 
              2 e^{-\left(x-v_{g_{+}}t\right)^2/2\sigma^2}
                e^{-\left(x-v_{g_{-}}t\right)^2/2\sigma^2}
         \cos\left[2k\left(x-v_{g_0}t\right)\right]\Big) \nonumber.
\end{eqnarray}
\end{widetext}
We see here that the moving Gaussian envelopes form a background for
the interference pattern which travels at a velocity $v_{g_0}$ in the inertial frame
(and thus appears stationary in the rotating frame as per Figs.~2(b,c,d)).
This is what gives rise to the apparent plateaus as measured in the
rotating frame.  The phase shift is still accumulating during the
collision, but that accumulation gives rise to a motion in the
interference fringes that exactly matches the rotation.
We also note that the interference pattern has a wavenumber of $2k$ which
we rely on in the phase-shift extraction algorithm.

\section{Ancillary - Videos}
\label{app:videos}

Three videos are available as Ancillary arXiv Files.
These are animations of 2-D calculations of our dispersion-based
Sagnac atom interferometers with initial states shown in Fig.~1(a)
corresponding to cases (i), (ii), and (iii).
These calculations are shown in the rotating frame,
with a fast rotation rate of $\Omega = 0.0150$, along with a weak
angular trap with $\omega_\phi=0.1$, chosen to enhance the
visualisation of the phase structures, especially for $t<0$.
This $\Omega$ is an order of magnitude faster than the calculations
shown in Figs.~3 and 4, however, it is the same as the rate
chosen for the 1-D calculations shown in Fig.~2.
Thus, the expected phase-shift at the first collision is
$\Delta(t) = 3\pi$, and thus $\Delta(t) = 6\pi$ at the second.
For video (i) we thus have the centre of the rotation
at $(x_\Omega,y_\Omega) = (0,0)$, and for (ii) and (iii) we have
$(x_\Omega,y_\Omega) = (-40,0)$.
To limit to a $240\times 240$ pixel video resolution, the axis
labels are not shown.  The range shown spans $x,y \in [-15,15]$
to enclose the ring of radius $R=10$ osc. units.

The videos display both parts of the wavefunction,
$\Psi(x,y,t) = A(x,y,t) \exp(i\varphi(x,y,t))$, at once.
Firstly the contours denote lines of equal probability density
($A^2$, chosen at $10^{-4}$, $10^{-2}$ and $10^{0}$),
and secondly coloured by the \textit{local phase} of the wavefunction
(i.e. $\varphi \in [0,2\pi)$).  The initial part of the videos show
the Gaussian initial states as they are evolved in imaginary time
to establish the true (rotating) ground state of the systems with
both of their $V_{\mathrm{ring}}$ and $V_{\mathrm{trap}}$ turned on
(note the lines of equal phase pointing towards $(x_\Omega,y_\Omega)$).
Once that has converged, we commence the real time calculation
(at $t=0$) by turning $V_{\mathrm{trap}}$ off, and the wavepackets
slowly disperse around the ring, with the animations
ending at $t=100$ osc. units.


%


\end{document}